\documentclass{ws-p8-50x6-00}

\begin{document}

\title{Orbital Magnetism in the Cuprates}

\author{Sudip Chakravarty,  Hae-Young Kee,  and Chetan Nayak}

\address{Department of Physics, University of California Los Angeles,\\  
Los Angeles CA 90095, USA\\ 
E-mail: sudip@physics.ucla.edu}

\maketitle

\abstracts{
The pseudogap phase of the cuprate superconductors is argued to be
characterized by a hidden broken symmetry of $d$-wave character in the
particle-hole channel that leads to staggered orbital
magnetism.  This proposal has many striking
phenomenological consequences, but  the most direct signature of this
order should be visible in the neutron scattering experiments. The
theoretical underpinning of these experiments is  discussed.}

\section{Introduction}
It has been proposed that the pseudogap state is a phase with a broken symmetry
and an order parameter,\cite{Chakrvarty1,Chakravarty2} not a state with a merely fluctuating
order.\cite{Lee} The distinction between the two is enormous.  The order parameter, which is due to a
particle-hole condensate with internal ``angular momentum" 2, will be termed the $d$-density wave (DDW). We
have predicted that the present day neutron scattering experiments have enough resolution to provide a direct
evidence of this order. And, indeed, there are some preliminary  neutron scattering measurements in the
bilayer YBa$_2$Cu$_3$O$_{6.6}$ that appear to have observed the theoretically predicted order
parameter.\cite{Mook,CKN} There are other experiments on YBa$_2$Cu$_3$O$_{6.5}$, which observe a magnetic
signal,\cite{Sidis} but appear to be from remnant spins left over from the undoped antiferromagnet. While the
experimental situation must be resolved, we shall focus merely on the theoretical predictions because a
precise  analysis is necessary to make sure that what is being observed is not an artefact. The fundamental
signature of this order parameter is an elastic Bragg peak centered at two-dimensional wave vector
${\bf Q}=(\pi/a,\pi/a)$, where $a$ is the lattice spacing of the CuO-lattice. Since the magnitude of the
staggred orbital fields is of order 10 G, the experiments are diificult.

The neutron scattering from DDW was addressed previously,\cite{Hsu} but the polarized neutron
scattering and the important question about the current form factors were not. The current form factors play
an important role in  polarized neutron measurements, without them one may be led to physically incorrect
results. Moreover, the crucial  connection to the pseudogap phase was not made, nor was it made clear
that such a phase truly survived beyond the large-$N$ mean field theory due to presumed gauge fluctuations.
There is also a proposal\cite{Varma} of a circulating current phase, which does not break translational
symmetry. The similarity with the DDW state is merely superficial. At the very least, the signature, on
robust symmetry grounds, is a neutron signal at
$Q=0$ which is fundamentally different. We will have nothing to say about this proposal, as it is not germane
to the present set of experiments.

\section{Order parameter}

The singlet DDW is defined by the order parameter $\Phi_{\mathbf Q}$ in terms of a
particle-hole condensate\cite{Nayak}
\begin{equation}
\langle {c^{\sigma\dagger}}({\bf k}+{\bf Q},t)
{c_\rho}({\bf k},t)\rangle
= i{\frac{\Phi_{\bf Q}}{2}}\,(\cos{k_x}a-\cos{k_y}a)\, {\delta^\sigma_\rho},
\end{equation}
where  $\sigma$, $\rho$ are spin indices.  The order parameter breaks time
reversal, parity, translation by a lattice spacing, and the rotation by $\pi/2$, although the product of any
two of these is preserved. Interestingly, DDW does not modulate charge at all, but it modulates currents. The
reason it is still called a density wave is because it is a particle-hole condensate, and
$d$-wave  refers to the internal form factor of the particle and the hole, which is $(\cos{k_x}a-\cos{k_y}a)$.
We expect that the magnetic field generated by the circulating currents should be proportional to the
pseudogap.\cite{Chakravarty2} A simple estimate results in a tiny field of order 10 G due to currents
circulating in a CuO plaquette as shown below:
\begin{figure}[htb]
\centerline{\includegraphics[scale=0.25]{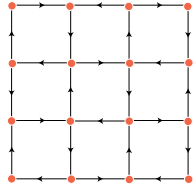}}
\caption{Circulating current pattern in the model where the current is carried by $\delta$-function
wires.}
\end{figure}

Although this simple model captures the correct symmetry, it is cruicially deficient in predicting
the correct neutron scattering intensities, as we shall see. It is {\sl a priori} clear that the flow of
currents cannot be along infinitesimally thin wires, but it must spread out, both because of 
finite sized atomic  orbitals and because of  electron-electron interactions.
One can model the more realistic current distribution\cite{CKN} by the form factors $\alpha({\bf q})$ and
$\beta({\bf q})$ in the Fourier transform of the current distribution $\langle {\bf j}({\bf
q})\rangle$without destroying the symmetry of the order parameter (${\bf q}
\cdot
\langle {\bf j}({\bf q})\rangle = 0$), where
\begin{eqnarray}
\langle{\bf j}({\bf q})\rangle &\propto& \Phi_{\bf Q}
\sum_{{\bf G}_{\parallel}}\delta_{{\bf q}_{\parallel},{\bf G}_{\parallel}}\,
\,f({\bf q})\nonumber\\& &  \times 
\left[\left(\alpha({\bf q})\frac{\hat{\bf x}}{q_x} -
\beta({\bf q})\frac{\hat{\bf y}}{q_y}\right) -
(\alpha({\bf q})-\beta({\bf q}))\frac{\hat{\bf z}}{q_z}\right].
\end{eqnarray}
Here, ${\bf q}_{\parallel}=(q_x,q_y)$,  $f({\bf q})=\sin(\frac{q_z d}{2} )$, $d$ is the separation
between the two CuO-planes within a bilayer complex, and
$\bf G$ is a reciprocal lattice vector.
The simplest choice is to take $\alpha({\bf q})$ and $\beta({\bf q})$ to be
dependent on
$q_z$ only, such that
$\alpha(q_z)-\beta(q_z)\sim q_z^2$,  as $ q_z\to 0$. The choice of this current density was discussed in
Ref.  5.
\section{Experimental detection of DDW}
The DDW is usually hard to detect  because
\begin{equation}
\sum_{\bf k}\langle {c^{\sigma\dagger}}({\bf k}+{\bf Q}) 
{c_\rho}({\bf k})\rangle\propto \sum_{\bf k} (\cos{k_x}a-\cos{k_y}a) = 0
\end{equation}
There is no net modulation of charge and spin that could be measured.
Contrast this with a $d$-wave superconductors (DSC) for which a similar momentum sum of the particle-particle
condensate also vanishes. Nonetheless, because of  broken gauge symmetry, Meissner effect follows,
irrespective of the pairing channel.
This is not possible for a DDW because the broken symmetry is in the Ising universality class, and experiments
seeking to uncover such order must (a) be sensitive to spatial variation of kinetic energy or currents, (b)
measure higher order correlations of the charge or spin density as in
2-magnon Raman scalletring, nuclear quadrupole resonance, etc.\cite{Nayak} One nice feature of DDW is that a
thermodynamic transition is possible in two dimensions due to the Ising nature of the order parameter.

Impurities will   mix $d$-wave order with the $s$-wave order proportional to the concentration of
impuritiues. This may be an indirect way to reveal the hidden broken symmetry, because the corresponding
$s$-wave order is none other than the conventional charge density wave (CDW).
Another possibility is spin-orbit coupling, which will mix spin density wave (SDW) with DDW. In mean field
theory,\cite{Zeyher} there is some evidence that  DDW may incommensurate for higher doping. If this is the
case, the charge order must inevitably be mixed in, an idea that can be experimentally probed by scanning
tunneling measurements inside the vortex core of the mixed DDW and DSC phase. The detailed nature of the
charge order appear to be highly nonuniversal, and certainly cannot be used as a diagnostic tool for theories
of pseudogap.

\section{Neutron scattering}

\subsection{Unpolarized neutron scattering}
For unpolarized neutron scattering, the differential scattering cross sections for Bragg scattering from an
ordered array of orbital currents
is given by\cite{CKN}
\begin{equation}
\left( \frac{d\sigma}{d\Omega}\right)\propto\frac{\left|\langle{\bf j}({\bf q})\rangle\right|^2}{q^2},
\end{equation}
where $\bf q$ now is the momentum transfer. We certainly do not know  the form factors $\alpha({\bf q})$ and
$\beta({\bf q})$ beyond their limiting forms, but a reasonable  guess can be made to  see the robust features
of the intensity modulation as a function of the momentum transfer. For this purpose, we have chosen
$\alpha({\bf q}) = f(q) e^{-(\frac{q_z}{q_0})^2}$, and $\beta({\bf q}) = f(q) e^{-(\frac{q_z}{q_1})^2}$. So,
the intensity is parametrized by a orthorhombicity parameter $\lambda({\bf q})=
e^{-[(\frac{q_z}{q_0})^2-(\frac{q_z}{q_1})^2]}$ and $\beta({\bf q})$. Note that all other factors, such as
the magnitude of the order parameter, drop out if we normalize with respect to a reference Bragg reflection
intensity. We then replace
$\beta({\bf q})$ by the well known Cu form factor,\cite{Tranquada} which is likely to be  an upper bound
because the orbital currents are more spread out in real space than atomic orbitals. We choose $q_0=2\pi/d$
and $q_1 =0.275 q_0$. We shall see that the unpolarized intensity is not sensitive to the choice of the
$\lambda$-parameter.

\begin{figure}[htb]
\centerline{\includegraphics[scale=0.5]{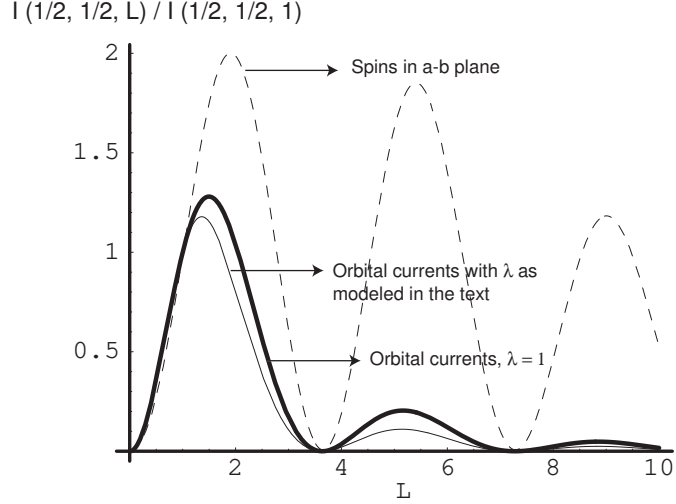}}
\caption{Normalized intensity, $I$, of Bragg reflections as a function of the perpendicular momentum transfer,
$q_z$. The shorthand notation is $I(H, K, L)\equiv I(2\pi H/a, 2\pi K/a, 2\pi L/c)$. The intensity from spins
lying in the $a$-$b$ plane is obtained from the form factor given in Ref. 11.}
\label{fig:intensity}
\end{figure}

There is no question that the unpolarized neutron scattering intensities are dramatically diffrent
for orbital currents in comparison to spins lying in the $a$-$b$ plane as shown in Fig.~\ref{fig:intensity}.
The scattering from spins pointing along the
$c$-direction can be easily calculated and is similar to the scattering from orbital currents. These two cases
can be distinguihed by going to higher order Bragg reflections such $(H, H, L)$, for $H > 1/2$, for example
$(3/2, 3/2, L)$. The results for orbital currents fall off very rapidly in contrast to spins pointing in the
$c$-direction.

\subsection{Polarized neutron scattering}

\begin{figure}[htb]
\centerline{\includegraphics[scale=0.25]{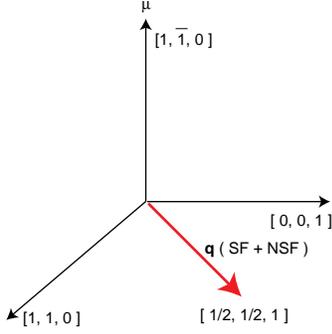}}
\caption{The vertical field geometry.}
\label{fig:coordinates}
\end{figure}

The Bragg scattering intensity for polarized neutrons from orbital currents is given
by
\begin{eqnarray}
\left( \frac{d\sigma}{d\Omega}\right)_{i\to f}
&\propto& \frac{1}{q^4} \left|\langle f|{\vec \mu}|i\rangle
\cdot{\bf q}\times\langle{\bf j}({\bf q})\rangle\,
\right|^2,
\end{eqnarray}
where $\vec\mu$ is the neutron spin, and $|i\rangle$ and $|f\rangle$ are the initial and the
final states of the neutron. It is easy to see that if the scattering vector is parallel to
the direction of polarization, the entire scattering is spin flip. This is useful to identify
magnetic scattering from non-magnetic scattering and the intensity ratios should follow the same
pattern as that of the unpolarized case. When the scattering vector $\bf q$ is perpendicular to
the polarization, there can be both spin-flip and non-spin-flip scattering. This geometry contains
additional information, which is the ratio of the spin-flip to non-spin-flip scattering. A common
experimental set up is to polarize neutron along along the $[1,\bar{1},0]$ direction as shown in
Fig.~\ref{fig:coordinates}, so that the new $x$-axis is along $[1, 1, 0]$, and the new $y$-axis is along $[0,
0, 1]$. 

Then a simple calculation shows that the ratio of the non-spin-flip to spin-flip scattering
intensity, $NSF/SF$, for Bragg reflection $(H,H,L)$ is 
\begin{equation}
\frac{NSF}{SF} = 2\,
{\left(\frac{\lambda(L)-1}{\lambda(L)+1}\right)^2}
\:\:{\left(\frac{Hc}{La}\right)^2}
\left[1+\frac{1}{2}{\left(\frac{La}{Hc}\right)^2}\right],
\end{equation}
which vanishes identically if there is no orthorhombicity, that is, $\lambda(L) =1$. This is a
nontrivial result for orbital current theory. 

For the choice of the parameters described earlier, 
corresponding to $\lambda(L=1)\approx 0.4$, this ratio for the reflection $(1/2, 1/2, 1)$ turns out to be
$\approx 1$. Thus, while modeling of the orthorhombicity of the current distribution had very little effect on
the unpolarized intensity, its effect on the polarized scattering is striking, turning the ratio $(NSF/SF)$
to order 1 as compared to  0.

\section{Concluding remarks}
Other possible evidence against spin order is the existence of a spin gap due to the Ising universality class
of the DDW order, and the small moment of order $2\times {10^{-2}} {\mu_B}$ corresponding to the field due to
circulating currents. We made an educated guess for the microscopic nature of the current distribution.
Clearly, we need a better understanding of it, although the ratio of the intensities as a function of
the momentum transfer for unpolarized scattering is very insensitive to the modeling of the current
distribution. Thus, we believe that our results are quite robust.

\section*{Acknowledgments}
We thank H. Mook and P. Dai for many discussions. This work was supported by a grant from
the National Science Foundation.

\end{document}